%% file: keychange.tex
\documentclass[sigconf,authorversion]{acmart}

\usepackage{caption}
\usepackage{subcaption}

\usepackage{placeins} 

\usepackage{stfloats}
\usepackage{float}

\usepackage{tabularx}
\usepackage{color, colortbl}

\usepackage{adjustbox}
\usepackage{array}
\newcolumntype{R}[2]{%
    >{\adjustbox{angle=#1,lap=\width-(#2)}\bgroup}%
    l%
    <{\egroup}%
}
\newcommand*\rot{\multicolumn{1}{R{45}{1em}}}

\usepackage{wasysym}

\usepackage[binary-units]{siunitx}
\definecolor{darkred}{rgb}{0.831, 0, 0.063}
\definecolor{sorange}{rgb}{0.95, 0.57, 0}
\colorlet{orange}{sorange}

\usepackage{listings}
\usepackage{xspace}

\usepackage[nolist]{acronym}

\acrodefplural{PoC}[PoCs]{Proofs of Concept}

\begin{acronym}
\acro{SDR}{Software-Defined Radio}
\acro{AGC}{Automatic Gain Control}
\acro{GCI}{Global Coexistence Interface}
\acro{ECI}{Enhanced Coexistence Interface}
\acro{AFH}{Adaptive Frequency Hopping}
\acro{FEM}{Front-End Module}
\acro{SP3T}{Single Pole, Triple Throw}
\acro{HCI}{Host Controller Interface}
\acro{A2DP}{Advanced Audio Distribution Profile}
\acro{SCO}{Synchronous Connection-Oriented}
\acro{GPIO}{General Purpose Input Output}
\acro{ASLR}{Address Space Layout Randomization}
\acro{LMP}{Link Management Protocol}
\acro{LCP}{Link Control Protocol}
\acro{LNA}{Low-Noise Amplifier}
\acro{EIR}{Extended Inquiry Response}
\acro{CRC}{Cyclic Redundancy Check}
\acro{RTOS}{Real-Time Operating System}
\acro{QEMU}{Quick Emulator}
\acro{UART}{Universal Asynchronous Receiver Transmitter}
\acro{MitM}{Machine-in-the-Middle}
\acro{MAC}{Media Access Control}
\acro{BLE}{Bluetooth Low Energy}
\acro{ACL}{Asynchronous Connection-Less}
\acro{DSP}{Digital Signal Processing}
\acro{SDIO}{Secure Digital Input Output}

\acro{BCS}{Bluetooth Core Scheduler}
\acro{ELF}{Executable and Linking Format}
\acro{GIAC}{Global Inquiry Access Code}
\acro{JSON}{JavaScript Object Notation}
\acro{L2CAP}{Logical Link Control and Adaptation Protocol}
\acro{LMP}{Link Management Protocol}
\acro{PTM}{Pseudo Terminal Master}
\acro{PTS}{Pseudo Terminal Slave}
\acro{QEMU}{Quick Emulator}
\acro{RCE}{Remote Code Execution}
\acro{RFU}{Reserved for Future Use}
\acro{SVC}{Supervisor Call}
\acro{UART}{Universal Asynchronous Receiver Transmitter}
\acro{WICED}{Wireless Internet Connectivity for Embedded Devices}
\acro{MMIO}{Memory Mapped Input/Output}
\acro{LE}{Bluetooth Low Energy}
\acro{IoT}{Internet of Things}
\acro{IDE}{Integrated Development Environment}
\acro{ARM}{Advanced RISC Machine}
\acro{LM}{Link Manager}

\acro{RTOS}{Real-Time Operating System}
\acro{DMA}{Direct Memory Access}
\acro{RXDMA}{Receive Direct Memory Access}
\acro{NVRAM}{Non-Volatile Random-Access Memory}
\acro{ROM}{Read-Only Memory}
\acro{Rx}{Receive}
\acro{Tx}{Transmit}
\acro{SPI}{Serial Peripheral Interface}
\acro{MSP}{Main Stack Pointer}
\acro{PSP}{Process Stack Pointer}
\acro{RF}{Radio Frequency}
\acro{BLOB}{Binary Large Object}
\acro{SCO}{Synchronous Connection Oriented}
\acro{UAF}{Use-After-Free}
\acro{FHS}{Frequency Hop Sync}
\acro{CRC}{Cyclic Redundancy Check}
\acro{PDU}{Protocol Data Unit}
\acro{NFC}{Near Field Communication}
\acro{JPEG}{Joint Photographic Experts Group}
\acro{LPE}{Local Privilege Escalation}
\acro{DEP}{Data Execution Prevention}
\acro{XN}{eXecute Never}
\acro{LCP}{Link Control Protocol}
\acro{GATT}{Generic Attribute}
\acro{PoC}{Proof of Concept}
\acro{EDR}{Enhanced Data Rate}
\acro{MWS}{Mobile Wireless Standards}
\acro{HAL}{Hardware Abstraction Layer}
\acro{LR}{Link Register}
\acro{eSIM}{embedded-SIM}
\acro{RSP}{Remote SIM Provisioning}
\acro{PC}{Program Counter}
\acro{MD}{More Data}
\acro{LLID}{Logical Link Identifier}
\acro{SFI}{Serial Flash Interface}
\acro{EEPROM}{Electrically Erasable Programmable Read-Only Memory}
\acro{TOFU}{Trust On First Use}
\acro{CVE}{Common Vulnerabilities and Exposure}
\acro{PCIe}{Peripheral Component Interconnect Express}
\acro{SECI}{Serial Enhanced Coexistence Interface}
\acro{HID}{Human Interface Device}
\acro{DoS}{Denial of Service}
\acro{PTA}{Packet Traffic Arbitration}
\acro{AWMA}{Alternating Wireless Medium Access}
\acro{AP}{Access Point}
\acro{ECDH}{Elliptic-Curve Diffie--Hellman}
\acro{BT}{Bluetooth Classic}
\acro{SSP}{Secure Simple Pairing}
\acro{SC}{Secure Connections}
\acro{MIC}{Message Integrity Check}

\end{acronym}

\usepackage[colorinlistoftodos,prependcaption,disable]{todonotes} 
\presetkeys%
    {todonotes}%
    {inline,backgroundcolor=orange}{}
    
\usepackage{blindtext}

\usepackage{mdframed}

\AtBeginDocument{%
  \providecommand\BibTeX{{%
    \normalfont B\kern-0.5em{\scshape i\kern-0.25em b}\kern-0.8em\TeX}}}

\copyrightyear{2021}
\acmYear{2021}
\setcopyright{acmlicensed}\acmConference[WiSec '21]{14th ACM Conference on Security and Privacy in Wireless and Mobile Networks}{June 28-July 2, 2021}{Abu Dhabi, United Arab Emirates}
\acmBooktitle{14th ACM Conference on Security and Privacy in Wireless and Mobile Networks (WiSec '21), June 28-July 2, 2021, Abu Dhabi, United Arab Emirates}
\acmPrice{15.00}
\acmDOI{10.1145/3448300.3467822}
\acmISBN{978-1-4503-8349-3/21/06}

\begin{document}

\title{Happy MitM -- Fun and Toys in Every Bluetooth Device}


\author{Jiska Classen}
\affiliation{%
  \institution{Secure Mobile Networking Lab, TU Darmstadt}
  \country{Germany}
}
\email{jclassen@seemoo.de}

\author{Matthias Hollick}
\affiliation{%
  \institution{Secure Mobile Networking Lab, TU Darmstadt}
  \country{Germany}
}
\email{mhollick@seemoo.de}

\renewcommand{\shortauthors}{Classen and Hollick}

\renewcommand{\shorttitle}{\includegraphics[height=0.8em]{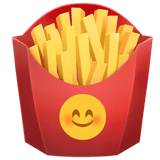} Happy MitM -- Fun and Toys in Every Bluetooth Device}

\input{sections/abstract.tex}




\maketitle

\input{sections/introduction.tex}

\input{sections/background.tex}
\input{sections/frida.tex}

\input{sections/vulns.tex}

\input{sections/conclusion.tex}

%

%
%

\begin{acks}
We thank Felix Rohrbach and Maximilian von Tschirschnitz for their feedback and discussion, and Adrian Dabrowski for shepherding our paper.
This work has been funded by the German Federal Ministry of Education and Research and the Hessen State Ministry for Higher Education, Research and the Arts within their joint support of the National Research Center for Applied Cybersecurity ATHENE.
\end{acks}

\balance

\bibliographystyle{ACM-Reference-Format}
\bibliography{bibfile}


\end{document}

%% file: sections/abstract.tex

\begin{abstract}

Bluetooth pairing establishes trust on first use between two devices
by creating a shared key.
Similar to certificate warnings in TLS, the Bluetooth specification
requires warning users upon issues with this key, because this can indicate
ongoing \ac{MitM} attacks.
This paper uncovers that none of the major Bluetooth stacks warns users, which violates the specification.
Clear warnings would protect users from recently published and potential future
security issues in Bluetooth authentication and encryption.
\end{abstract}

%% file: sections/introduction.tex
\section{Introduction}

\begin{figure}[!t]
\centering
%
%
%
%
%
%
	\begin{tikzpicture}[minimum height=0.55cm, scale=0.8, every node/.style={scale=0.8}, node distance=0.7cm]

    \node[inner sep=0pt] (iphone) at (-4,0)
    {\includegraphics[height=2cm]{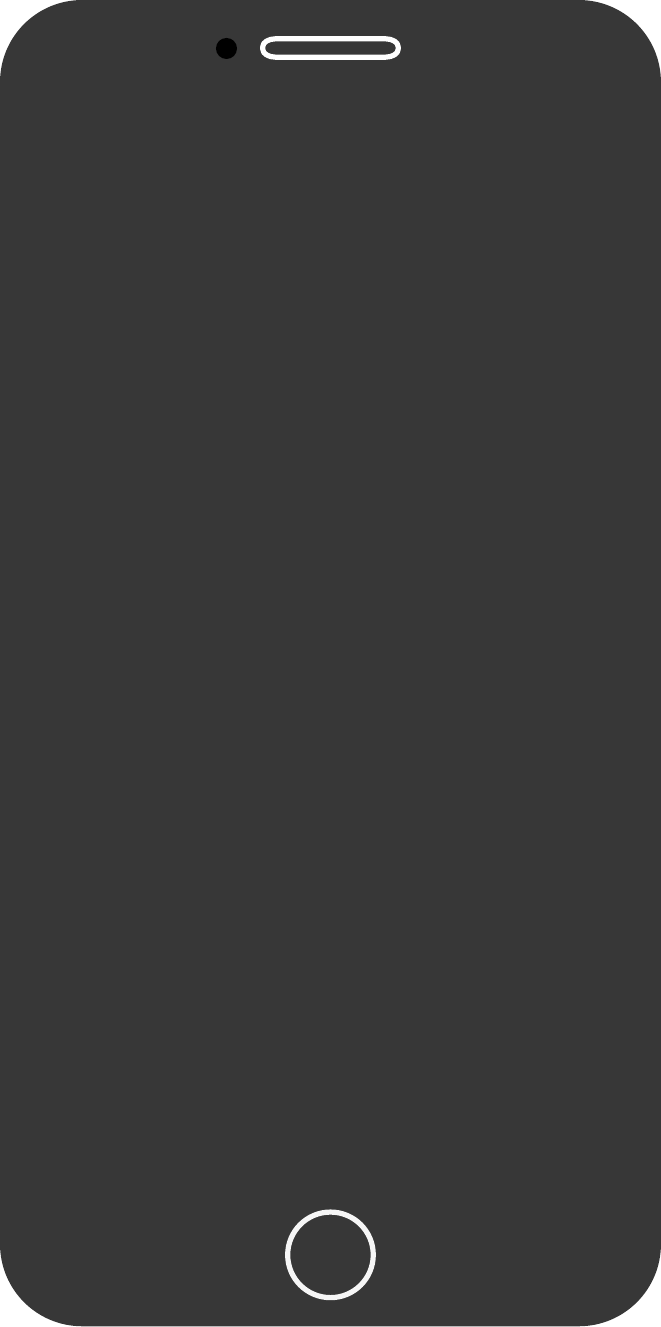}}; 
    \node[inner sep=0pt] (iphonex) at (-4,0.05)
    {\includegraphics[height=1.6cm]{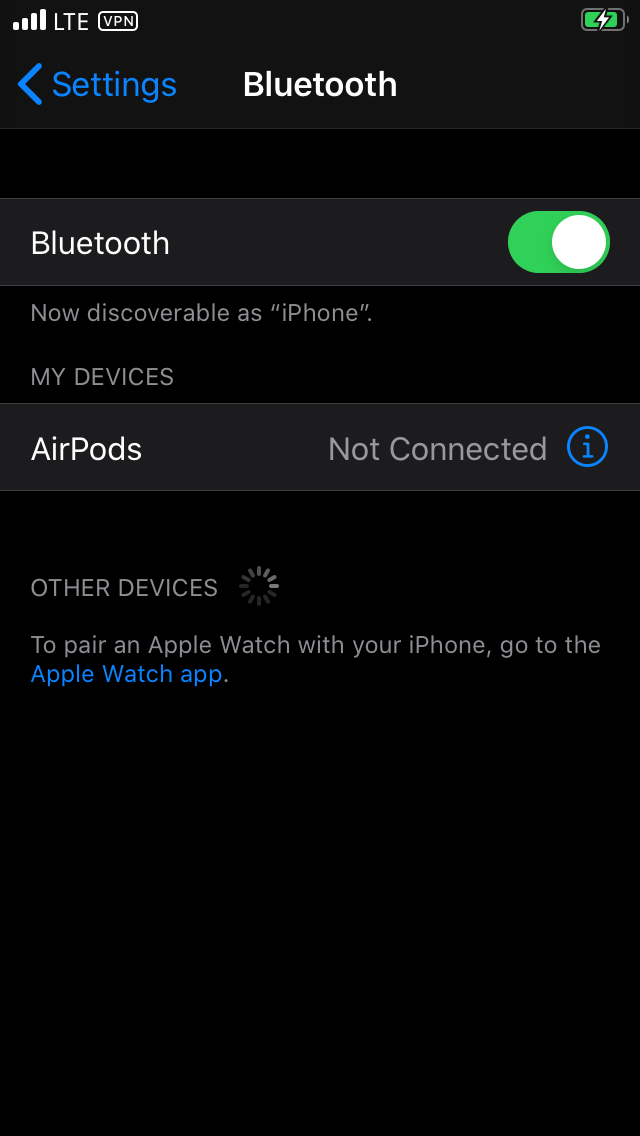}};  
    \node[color=darkred] (iphonetxt) at (-3.1, -1) {$K_{AM}$};

    \node[inner sep=0pt] (sdr) at (0,0.2)
    {\includegraphics[height=1cm]{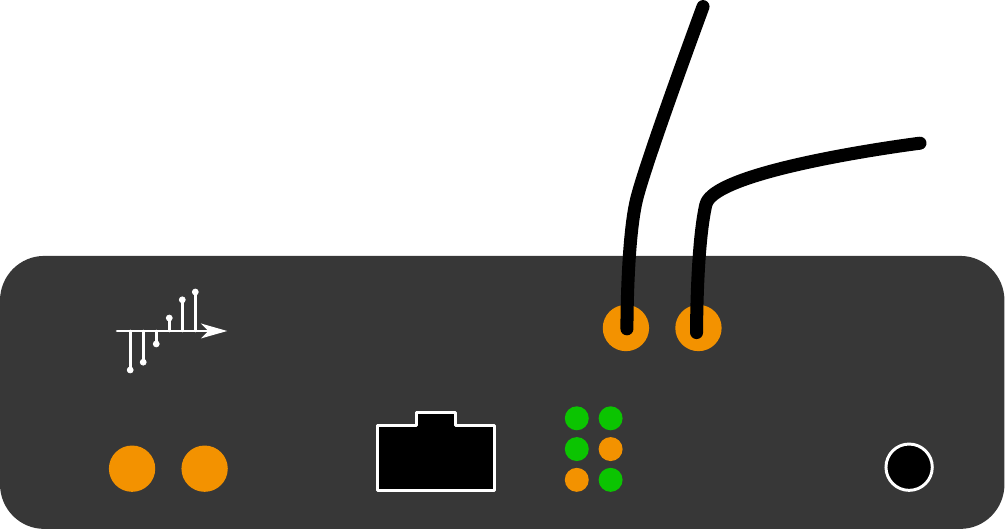}};    
    \node[color=darkred] (sdrtxt) at (0, -1) {$K_{AM}$, $K_{MB}$};	
   	
    \node[inner sep=0pt] (headset) at (4,0)
    {\includegraphics[height=1cm]{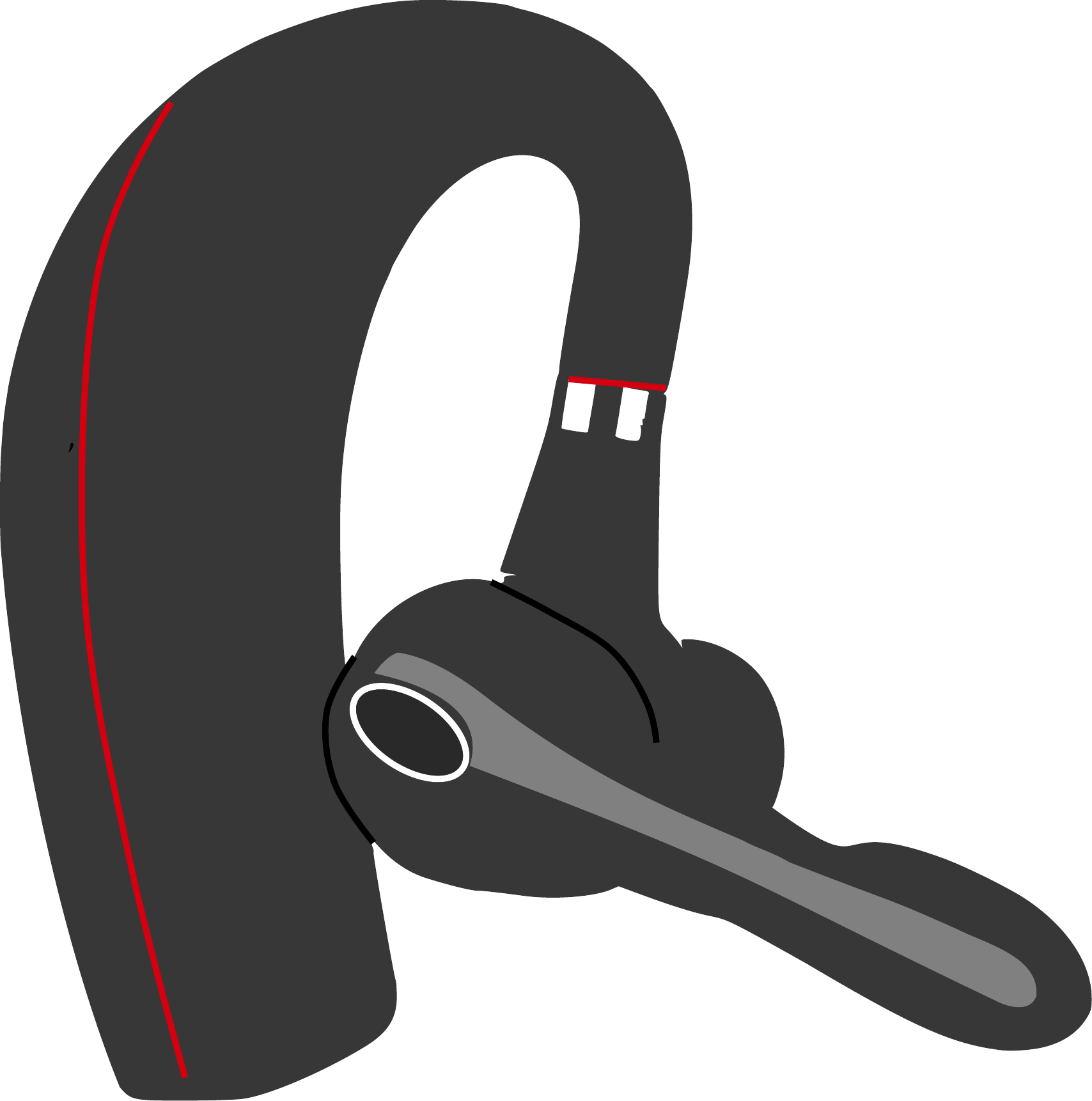}}; 
    \node[color=darkred] (headsettxt) at (4, -1) {$K_{MB}$};
    
    \draw[<->, color=darkred, dashed] (-3.35,0) -- (-1.1,0);
    \draw[<->, color=darkred, dashed] (1.1,0) -- (3.3,0);

	\end{tikzpicture}
\vspace{-0.5em} 
\caption{Most MitM attacks result in two separate keys.}
\label{fig:mitmactive}
\vspace{-1em} 
\end{figure}
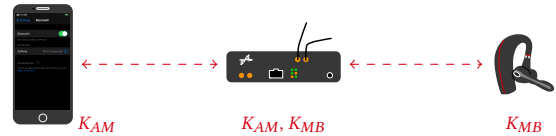

Attacks on Bluetooth pairing often lead to two separate keys, as shown in \autoref{fig:mitmactive}.
Such attacks are either enabled by vulnerabilities in the specification and implementation~\cite{methodconfusion, rng, reflection}
or the insecure \emph{Just Works} mode used by most IoT devices and headsets~\cite[p. 985]{bt52}.
In practice, an attacker faces the following barriers:

\begin{enumerate}
\item Presence during the initial pairing or forcing a new pairing.
\item Presence in all future connections to re-encrypt traffic.
\end{enumerate}

In other network protocols, such as TLS, continuous presence can be achieved by placing a \ac{MitM} on, e.g.,
a router close to the target. Bluetooth is used on devices that move and have varying signal strength.
\textbf{A permanently successful attacker must be omnipresent and immediately reply with a strong signal to all connection attempts.} If the attacker only fails once---which is very likely in a mobile environment---devices under attack would use incompatible keys,
resulting in an authentication or encryption failure. According to the Bluetooth 5.2 specification,
the user shall be notified of security failures~\cite[p. 1314]{bt52}.
\textbf{We find that all major Bluetooth stacks skip warning the user, thereby violating the
 specification.}
Warnings are independent from the underlying pairing method and technology, since pairing and connection
dialogues are implemented on top. We test user interfaces on a large variety of devices, ranging from Bluetooth 2.1 + EDR to 5.2,
including \ac{BT} and \ac{BLE} as well as the pairing extensions \emph{Google Fast Pair} and \emph{Apple MagicPairing}~\cite{fastpair,magicpairing}.
More precisely, the following platforms are affected:

\begin{itemize}
\item Both tested \emph{Android} flavors (\emph{Google} and \emph{Samsung}) do not indicate authentication failures to the remote device.
\item \emph{Google Android} further silently removes the pairing, which opens the door for enforcing new pairings.
\item \emph{iOS}, \emph{Samsung Android}, and \emph{Windows} display a message that they could not connect without explaining why, and \emph{macOS} as well as \emph{Ubuntu Gnome} indicate a failed connection via user interface button colors. The original key stays valid.
\item Various gadgets do not indicate any error and keys stay valid.
\end{itemize}

%
%

We demonstrate stack and user interface failures against both \ac{BLE} and \ac{BT}, using F\reflectbox{R}IDA~\cite{frida} to dynamically hook the \emph{iOS} and \emph{Android} Bluetooth stacks and substitute keys in \ac{HCI} commands. In contrast to existing tools, this allows conditional interaction with the Bluetooth stack by altering commands and events. These scripts are now part of the \emph{InternalBlue} framework~\cite{mantz2019internalblue}, as they will be valuable for further Bluetooth-related research.

This paper is structured as follows.
\autoref{sec:background} explains Bluetooth pairing and security fundamentals.
Then, \autoref{sec:frida} continues with details on how to hook into Bluetooth stacks to test for security issues.
All identified vulnerabilities are detailed in \autoref{sec:vulns}.
\autoref{sec:conclusion} concludes this paper.

%% file: sections/background.tex

\section{Background}
\label{sec:background}

The following section explains Bluetooth pairing basics, recent attacks,
and expected failures 
in case of \ac{MitM} presence.

\subsection{Bluetooth Pairing Modes}

Bluetooth pairing modes and warnings in user interfaces are separate components
on all stacks researched in this paper. However, security vulnerabilities in
Bluetooth pairing or encryption enable \ac{MitM} attacks, and, thus, motivate
clear warnings in user interfaces.

The early \ac{BT} versions had a very flawed pairing, now termed
\emph{Legacy Pairing}, since it still is implemented for backwards
compatibility~\cite{2005:shaked}. \ac{BT} 2.1 introduced \ac{SSP}, which was formally verified~\cite{chang2007formal}.
Despite this verification, the specification is unclear about
certain aspects of \ac{SSP} and the follow-up encryption, resulting in various
practical attacks~\cite{hypponen2007nino, 2018:biham, methodconfusion, antonioli20bias, knob, reflection}.

With the Bluetooth specification version 4.0, \ac{BLE} was introduced, featuring low-energy connections for IoT
gadgets and medical devices. Instead of using exactly
the same pairing mechanism, a lightweight pairing was added, which is fundamentally broken
and now called \emph{LE Legacy Pairing}~\cite{ryan2013bluetooth}.
Newer versions use \ac{SC}, which are very similar to \ac{SSP}.
Thus, attacks on \ac{SSP} typically also apply to \ac{SC}. Even worse, the key format in \ac{BT}
and \ac{BLE} devices is rather similar, and there is a cross-transport key derivation for both protocol variants, which 
also is vulnerable~\cite{antonioli2020blurtooth}.

\begin{table}[!b]
\renewcommand{\arraystretch}{1.3}

\newcolumntype{P}[1]{>{\centering\arraybackslash}p{#1}}

\caption{Authentication failure actions as defined by the Bluetooth 5.2 specification~\cite[p. 1314]{bt52}.}
\label{tab:authfail}
\vspace{-1em} 

\centering
\footnotesize
\begin{tabular}{l P{0.7cm} p{5cm}}
\textbf{Link Key Type} & \textbf{Bonded} & \textbf{Action} \\
\hline
Combination & $\times$ & \emph{\underline{Option 1:}} Automatically initiate pairing. \vspace*{0.5em}\newline \emph{\underline{Option 2, recommended:}} Notify user and ask if pairing is ok.\\
\rowcolor{gray!20}
Combination & \checkmark & Notify user of security failure.\\
Unauthenticated & $\times$ & \emph{\underline{Option 1, recommended:}} Automatically initiate SSP.\vspace*{0.5em}\newline \emph{\underline{Option 2:}} Notify user and ask if SSP is ok.\\
\rowcolor{gray!20}
Unauthenticated & \checkmark & Notify user of security failure.\\
Authenticated & $\times$ & \emph{\underline{Option 1:}} Automatically initiate SSP.\vspace*{0.5em}\newline \emph{\underline{Option 2, recommended:}} Notify user and ask if SSP is ok.\\
\rowcolor{gray!20}
Authenticated & \checkmark & Notify user of security failure.\\
\hline
\end{tabular}

\end{table}

From a user perspective, the underlying pairing mechanism is opaque.
No matter if \ac{SSP}, \ac{SC}, or \emph{LE Legacy Pairing} is used,
they all feature a \emph{Just Works} mode, which is vulnerable to active \ac{MitM}
attacks~\cite[p. 985]{bt52}.
They also have a \emph{Numeric Comparison} and \emph{Passkey Entry} mode, which can
prevent active \ac{MitM} attacks---assuming that all known vulnerabilities are fixed.
The \emph{Out Of Band} mode is not affected by these attacks, since it uses 
a non-Bluetooth channel for exchanging keys. Its vendor-specific implementation may or may not have flaws,
and attacks on stages after pairing such as session key entropy reduction~\cite{knob} apply either way.

This attack history without possibility to secure the \emph{Just Works} mode led to
manufacturers implementing independent pairing solutions. \emph{Apple} uses so-called
\emph{MagicPairing}, which is undocumented but has been reverse-engineered~\cite{magicpairing}.
\emph{Google} has a similar protocol called \emph{Fast Pair}~\cite{fastpair}.
Both protocols bind Bluetooth keys to cloud accounts and share them across devices.
Users only pair a headset once and then can access it via all devices logged into
the same cloud account. This is not only more convenient but also reduces the amount of
pairing attempts during which \acp{MitM} could be present.

\subsection{Attacks Resulting in Different Keys}

The majority of attacks on Bluetooth pairing results in a setup with different
keys. 
This applies to all devices using \emph{Just Works} mode as well as attacks
downgrading a pairing to this mode~\cite{hypponen2007nino}, mixing paring
modes~\cite{methodconfusion}, and reflecting messages~\cite{reflection}.
Besides direct attacks on protocols, implementation details such as a weak random number generator
can also enable \ac{MitM} attacks~\cite{rng}.



Even a successful attack on the pairing requires a \ac{MitM} to be omnipresent during
all follow-up connections.
\textbf{Otherwise, inconsistent keys lead to authentication and encryption errors.}
These errors can and should be used to detect attack attempts according to the Bluetooth
specification, as described in the following.

\subsubsection{Expected Authentication Failure Behavior}

If authentication fails, the host terminates the connection~\cite[p. 1959]{bt52}.
Moreover, the user should---in most cases---be warned~\cite[p. 1314]{bt52}.
The Bluetooth specification has a way more sophisticated decision process, as shown in \autoref{tab:authfail}.
A \emph{Combination} key is meant for \ac{BT} and \ac{BLE}, thus, initiating either \ac{SSP} or \ac{SC} upon a failure is valid.
\emph{Unauthenticated} keys are the result of \emph{Just Works} mode pairing~\cite[p. 1306]{bt52}, as it does not protect against \ac{MitM}.
In contrast, an \emph{Authenticated} key requires \emph{Numeric Comparison}, \emph{Passkey Entry}, or \emph{Out Of Band} pairing.
\emph{Bonded} devices permanently store the keys established during the initial pairing to create a trusted relationship.

If a key is \emph{Unauthenticated} and does not protect against active \ac{MitM}, the recommended option is to automatically
initiate a new \ac{SSP} for non-bonded devices. \textbf{This violates the trust on first use concept and \ac{MitM} attacks can successfully be
launched without any user interaction.} This is also the default option for non-bonded \emph{Combination} keys as well as the alternative
option for non-bonded \emph{Authenticated} keys.

Authentication might legitimately fail if one of the devices deleted the according key. 
This requires the user to manually reset a device, meaning that the user is aware and can act accordingly.

\subsubsection{Expected Encryption Failure Behavior}

\ac{BLE} devices activate encryption using the \path{LE_Enable_Encryption} \ac{HCI}
command~\cite[p. 2322]{bt52}. Then, the \ac{BLE} link layer tries to initiate encryption.
After finishing this step, an \emph{Encryption Mode Change} event is sent to the host,
indicating if encryption is on or still off.
When replacing \ac{BLE} keys, this results in encryption being reported as off~\cite[p. 2299]{bt52}.
The specification considers encryption failures in \ac{BLE} in case that the
remote device does not support encryption~\cite[p. 3141]{bt52}, and there is no differentiation
to having an invalid key. Thus, for \ac{BLE} encryption failures, the overall behavior is not
specified in detail.


%% file: sections/frida.tex

\section{Bluetooth Stack Test Framework}
\label{sec:frida}

As explained in the following, over-the-air \ac{BT} \ac{MitM} setups are still
rather expensive. To facilitate testing \ac{BT} and \ac{BLE} security, we instead
dynamically hook into \ac{HCI}.

\subsection{Over-the-Air Setups}
As of now, \acf{BLE} \ac{MitM} setups can be realized with the \texttt{btlejack}
toolsuite and three \emph{Micro:Bits}~\cite{btlejack}. This makes a \ac{BLE} setup as cheap as
USD 45.

There is still no
similar open-source tool for \acf{BT}, which has a more complex modulation scheme making
eavesdropping harder. Just in 2020, the first full-band \ac{BT} sniffer
for \acp{SDR} and more recent Bluetooth specifications was released~\cite{francescosniffer}. This sniffer supports
synchronization, dewhitening, decoding, as well as an algorithm to deanonymize
addresses.
This setup requires two USRP B210, totaling in approximately USD 3000.
For an active \ac{MitM} setup, even this advanced testing is insufficient,
since it does not support sending packets.

Nonetheless, it is reasonable to assume that strong attackers have a working \ac{MitM} setup.
From a technological standpoint a \ac{MitM} only needs a recent \ac{SDR} with \SI{80}{\mega\hertz} bandwidth---and move significant parts
of the implementation into an FPGA to fulfill real-time requirements.
Given the current progress of \acp{SDR}, it is to be expected that there will be
affordable \ac{BT} \ac{MitM} setups soon, meaning that devices should be proactively
secured against them.

\subsection{Hooking into Bluetooth Stacks}

%

Analyzing device behavior upon authentication failures due to changed keys
only requires changing the key locally on one of the paired devices. By changing
the key back and forth, it can even be tested if the original key is still trusted after
in between authentication failures.
There are two options to change a key:
\begin{enumerate}
\item Change the key within the host's file system.
\item Substitute the key within \ac{HCI} commands.
\end{enumerate}

The first option usually requires to restart the Bluetooth daemon. Moreover,
changing the key on the file system while Bluetooth is still running might
corrupt the local key database. 

Replacing key information within \ac{HCI} commands, the second option, is more flexible. Most
Bluetooth chips only have a ROM and need to request the key either on first
usage or whenever they establish a connection, depending on the implementation.
\emph{Apple} \emph{MagicPairing} and \emph{Google Fast Pair}
use this property~\cite{magicpairing, fastpair}:
They manage keys bound to cloud accounts separately, 
but still use the encryption mechanisms provided by the
Bluetooth chip. Thus, substituting keys as they are requested even allows
testing vendor-specific protocol additions. 
Key change behavior can be tested by replacing the \path{HCI_Link_Key_Request_Reply}
command for \ac{BT}~\cite[p. 720]{bt52}, respectively the \path{LE_Enable_Encryption} command for 
\ac{BLE}~\cite[p. 2322]{bt52}.

Note that it is not possible to use the original version of \emph{Internal\-Blue}~\cite{mantz2019internalblue}
for injecting different keys into the controller, because the controller would ask
the host for a key belonging to a specific address, expecting a single response.
Prior to the hooks we published along with this paper,
\emph{InternalBlue} only supported injecting commands but could not replace contents
of existing commands.

The \emph{Linux BlueZ}~\cite{bluez}
stack stores connection properties in separate files per connection, making
the first option suitable. The \ac{HCI}-based alternative
on \emph{Android} and \emph{iOS} enables us to also test vendor-specific additions,
which are not implemented on \emph{Linux}.

\subsubsection{Linux}

On \emph{Linux}, keys for \ac{BT} and \ac{BLE} connections are stored in
\path{/var/lib/bluetooth/mac1/mac2/info}, where \path{mac1} represents the
device address of the controller and \path{mac2} represents the
address of the paired device. After replacing the keys, the Bluetooth daemon
must be restarted to refresh information from these files.

Hooking \ac{HCI} in user space is infeasible on \emph{Linux}.
The Bluetooth daemon only parses a management
protocol, which further abstracts \ac{HCI}. The \emph{Linux}
kernel parses this custom management protocol and translates it into \ac{HCI}
commands and events. Having \ac{HCI} functionality in the kernel space
does not allow F\reflectbox{R}IDA-based hooks. 
If needed, more flexible hooks could be achieved by modifying the \ac{HCI} layer in
\path{/net/bluetooth} in the kernel source~\cite{linuxkernelbt}.

\subsubsection{Android}

The \emph{Android Fluoride} Bluetooth stack is open-source. The \ac{HCI} implementation
is contained in the file \path{system/bt/hci/src/hci_layer.cc}~\cite{androidhci}.
Commands to the controller are sent using the \path{transmit_command} function, and
events from the controller pass the \path{filter_incoming_event} function.
After compilation, these functions end up in the \path{libbluetooth.so} binary.

Despite having source code access, recompiling the stack to swap the key would not
be a flexible solution for more generic \ac{HCI} analysis. Instead, we hook the stack
using F\reflectbox{R}IDA~\cite{frida} on a rooted \emph{Samsung Galaxy Note20 5G}
with the January 2021 patch level. On this device, \path{libbluetooth.so} does not
contain symbols. Thus, we locate the relative address of \path{transmit_command} manually
using \emph{IDA Pro 7.5}~\cite{ida}. Based on this initial \emph{IDA} database, the address of this function
can be found automatically using \emph{BinDiff}~\cite{bindiff} if \path{libbluetooth.so}
was compiled for the same architecture.

\subsubsection{iOS}

The closed-source \emph{iOS} Bluetooth stack can be reverse-engineered using
debug strings, which even contain some of the original function names.
The part of the Bluetooth stack responsible for \ac{HCI} is contained in the Bluetooth
daemon \texttt{bluetoothd} itself instead of using a separate library.
Since functions implementing \ac{HCI} play an important role, they all contain debug
strings. Every command passes the function \path{OI_HciIfc_CopyPayload}, and every
incoming event is processed by \path{OI_HCIIfc_DataReceived}.

Similar to the \emph{Android} setup, we jailbreak an \emph{iPhone 8} on
\emph{iOS 14.4} with \emph{checkra1n}~\cite{checkra1n}, and use a F\reflectbox{R}IDA script to hook \ac{HCI}. The function
names required for the functionality we need stayed the same since various \emph{iOS}
releases, at least since \emph{iOS 13.5}, meaning that it should be easy to port the
hook to future \emph{iOS} versions.


%% file: sections/vulns.tex
\section{Vulnerable Stacks}
\label{sec:vulns}

We find that under some circumstances controllers do not indicate 
authentication failures (see \autoref{ssec:vulnlmp}).
Even if the controller issues such a failure, the user is not
notified on all tested stacks,
but they show varying behavior as shown in \autoref{tab:authfailvuln} (see \autoref{ssec:vulngoogle}--\ref{ssec:vulnperipherals}).
While testing popular peripherals, we discover that they use dangerously outdated Bluetooth versions (see \autoref{ssec:vulnoutdated}).
We disclosed all issues to the vendors (see \autoref{ssec:disclosure}).

\begin{table*}[tp]
\renewcommand{\arraystretch}{1.3}

\newcommand{\colorind}{\includegraphics[width=0.7em]{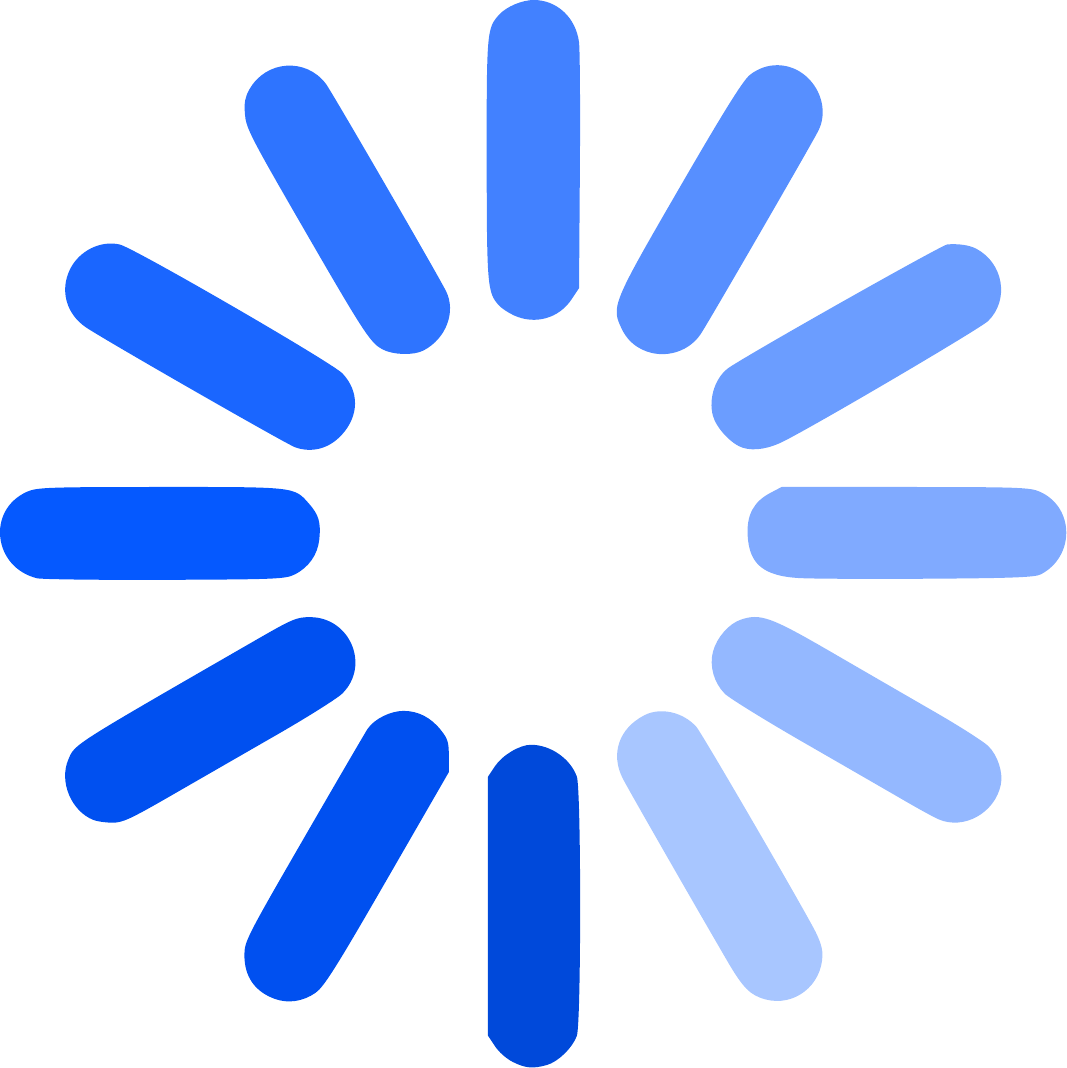}\hspace*{0.15em}}
\newcommand{\abortind}{\includegraphics[width=0.7em]{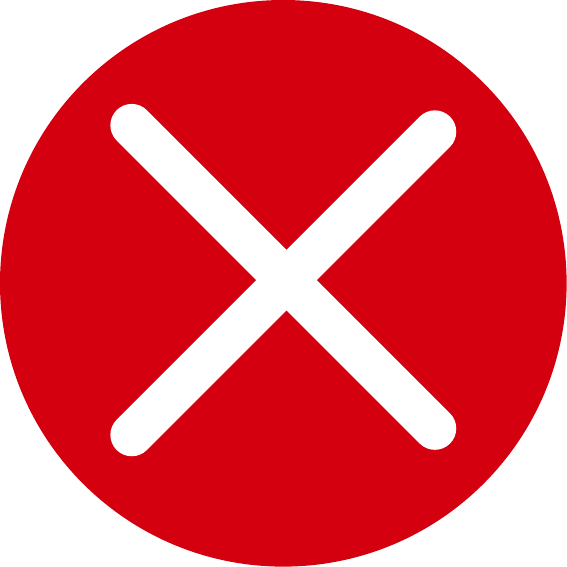}\hspace*{0.15em}}
\newcommand{\trashind}{\includegraphics[width=0.7em]{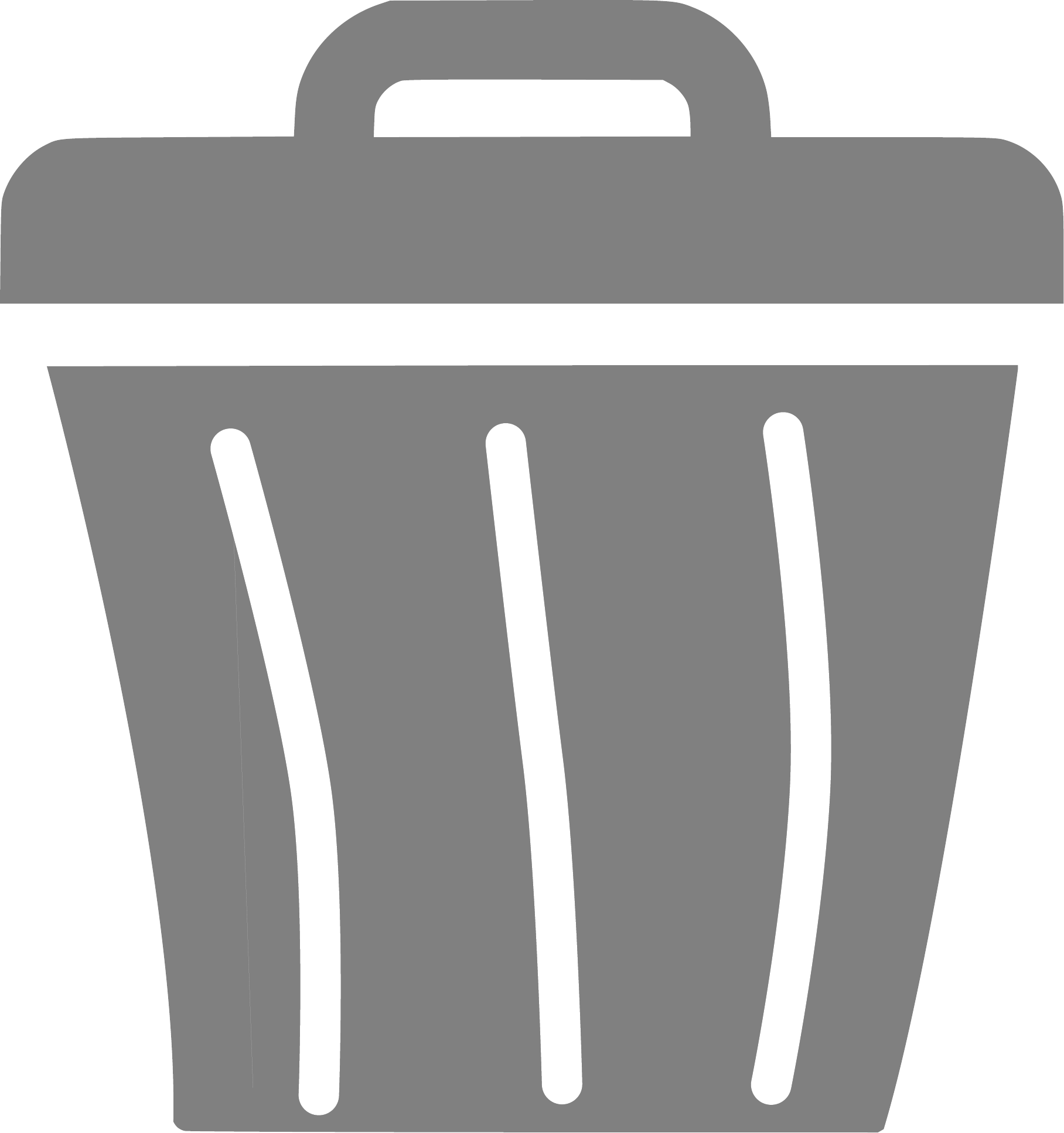}\hspace*{0.15em}}
\newcommand{\noind}{\textcolor{gray}{\CIRCLE}}

\caption{User notifications upon authentication and encryption failures due to invalid keys.}
\label{tab:authfailvuln}
\vspace{-1em} 
\centering
\footnotesize
\hspace*{-6.4cm}\begin{tabular}{lccccccccccccc}


\textbf{Key Fault Injector}
& \rot{\textbf{Device Under Test}}
& \rot{MacBook \textcolor{gray}{(macOS 11.2.1, BT 5)}} &
\rot{Ubuntu+Gnome \textcolor{gray}{(Mar 21, BT 4.0)}} & 
\rot{Windows 10 \textcolor{gray}{(Beta 21327, BT 4.0)}} & 
\rot{iPhone 8 \textcolor{gray}{(iOS 14.4, BT 5)}} & 
\rot{iPhone 12 \textcolor{gray}{(iOS 14.5 Beta, BT 5.2)}} & 
\rot{Galaxy Note 20 5G \textcolor{gray}{(Jan 21, BT 5)}} & 
\rot{Google Pixel 5 \textcolor{gray}{(Mar 21, BT 5.2)}} & 
\rot{AirPods \textcolor{gray}{(BLE 5)}} & 
\rot{Bose QC35 II \textcolor{gray}{(BT 4.2)}} & 
\rot{MagicKeyboard \textcolor{gray}{(BT 3.0 + HS)}} & 
\rot{Mini Keyboard \textcolor{gray}{(BT 2.1 + EDR)}} &
\rot{Xiaomi MI Band 2 \textcolor{gray}{(BLE 4.1)}} \\
\hline 

Invalid Key Effect                 & & \colorind & \colorind & \abortind & \abortind & \abortind & \abortind & \trashind &
\noind     & \noind & \noind  & \noind & \noind \\
\hline

iPhone 8                 & &  &  &  &  &  &  &  &
     &  &   &  & $\times$\\
Samsung Galaxy Note 20 5G  & &  &  &  &  &  &  &  &
$\times$    &  &   &  &  \\
Ubuntu with Gnome          & &  &  &    &  &  &  &  &
$\times$    &  &   &  & $\times$\\
\hline

\end{tabular}

\captionsetup{%
   labelsep=newline,
   justification=raggedright,
   labelfont=bf,
   singlelinecheck=off,
   width=1.0\textwidth 
}

\vspace{-1.85cm}\caption*{\normalfont\footnotesize{\hspace*{12cm}\noind ~No indication, key stays valid.\\
\hspace*{12cm}\colorind ~Button color or symbol indication, key stays valid.\\
\hspace*{12cm}\abortind ~Connection error text message, key stays valid.\\
\hspace*{12cm}\trashind ~Pairing is removed without user notification.\\ 
~ \\
\hspace*{12cm}$\times$ \hspace*{-0.05em}~Connection type not supported.\\
}}
\vspace{-1em} 
\end{table*}

\subsection{Bluetooth Controllers and LMP}
\label{ssec:vulnlmp}

While running the attacks, we capture traces using the \emph{Apple Packet\-Logger}
for \emph{iOS} and \emph{macOS}, \emph{Wireshark} on \emph{Linux}, and extract the
\path{btsnoop_hci.log} from \emph{Android} devices. All traces show that initiating \ac{BT}
controllers issue an \emph{Authentication Failure} event and \ac{BLE} controllers
issue an \emph{Encryption Failure} event. \textbf{This means that the underlying Bluetooth chips
are specification-compliant.}


Upon an authentication failure, the host shall terminate the connection~\cite[p. 1959]{bt52}.
The packet traces indicate that the Bluetooth hosts indeed follow this
procedure---none of the connections in \autoref{tab:authfailvuln} persisted after an
authentication failure.
Interestingly, while the \ac{BT} initiator always receives an \emph{Authentication Failure} via \ac{HCI}, some
responders only see a disconnect event with the reason \emph{Remote User Terminated Connection}.
We further analyze this by connecting the \emph{Samsung Galaxy Note20 5G} to a \emph{Google Nexus 5}.
The \emph{Nexus 5} is rather old, but supports \ac{LMP} sniffing via \emph{InternalBlue}~\cite{mantz2019internalblue}
and features \ac{SSP} with \ac{BT} 4.1. 

During the secure authentication phase, 
the initiator and responder can both end the connection with an \path{LMP_DETACH} packet containing the error code \emph{Authentication Failure}~\cite[p. 622]{bt52}.
The \ac{LMP} description of handling authentication errors is not in line with the \ac{HCI} part of the specification.
\ac{LMP} is only accessible by the controller and not the host, but if the host terminates the connection, it needs to issue
an \ac{HCI} command towards the controller. If authentication fails on the initiator, 
the controller correctly issues an \ac{HCI} event to the host indicating an \emph{Authentication Failure}. Then, the host sends an
\ac{HCI} command to terminate the connection, falsely using the error code \emph{Remote User Terminated Connection}.
Thus, the follow-up \path{LMP_DETACH} packet falsely contains the same error code.
\textbf{As a result, the disconnect event on the responder does not indicate an authentication failure.}
This affects at least devices with \emph{Android} versions \emph{6.0.1}--\emph{11}.

%

In addition, the host shall notify the user of a security issue upon an \emph{Authentication Failure}~\cite[p. 1314]{bt52}.
When looking at popular user interfaces, we avoid that \emph{Android} will not notify
the responder of authentication failures by testing both ends in the initiator role.

\subsection{Android (Google)}
\label{ssec:vulngoogle}

The \emph{Pixel 5} on the March 2021 patch level shows the most unexpected behavior.
Upon an authentication failure, the pairing entry is deleted. This happens without an
additional explanation---the user taps the device they want to connect to and next it
disappears from the list of paired devices.
Under certain circumstances, deleting keys is legitimate. Following \autoref{tab:authfail},
the specification states:

\begin{mdframed}[innerleftmargin=2em,linecolor=white]
\emph{``Non-bonded authenticated or unauthenticated link keys may be considered
disposable by either device and may be deleted at any time.''}~\cite[p. 1314]{bt52}
\end{mdframed}

The grammar in the previous sentence is unclear, but we assume that non-bonded
link keys, no matter if authenticated or not, can be deleted.
All devices under test were paired with the \emph{Pixel 5} using the \emph{Numeric Comparison}
method. The \ac{BT} link keys were authenticated and bonded.
\textbf{Deleting bonded keys
enables \ac{MitM} attackers to remove existing pairings with minimal user interaction---and then launch an attack on
the initial pairing.}
The \emph{Nexus 5} shows the same behavior on \emph{Android 6.0.1}, meaning that this
issue is consistent throughout \emph{Google}-flavored \emph{Android} versions.

\begin{figure}[!b]
\centering

\begin{subfigure}{1.0\columnwidth}
\center
	\includegraphics[width=0.45\columnwidth]{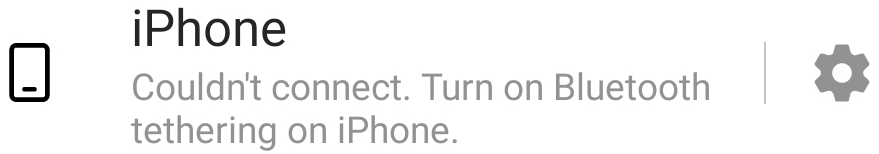}
    \caption{Samsung-flavored Android error message when the tethering AP changes its key.}
    \label{fig:sandroid_keychange}
\end{subfigure}


\begin{subfigure}{1.0\columnwidth}
\center
	\includegraphics[width=0.35\columnwidth]{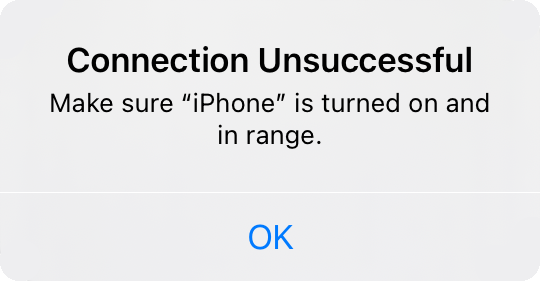}
    \caption{iOS error message when the tethering AP changes its key.}
    \label{fig:ios_keychange}
\end{subfigure}


\begin{subfigure}{1.0\columnwidth}
\center
	\includegraphics[scale=0.5]{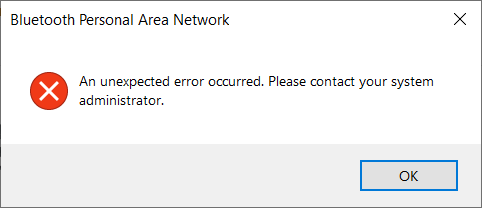}
    \caption{Windows error message when the tethering AP changes its key.}
    \label{fig:win_keychange}
\end{subfigure}


\begin{subfigure}{1.0\columnwidth}
\center
	\includegraphics[scale=0.5]{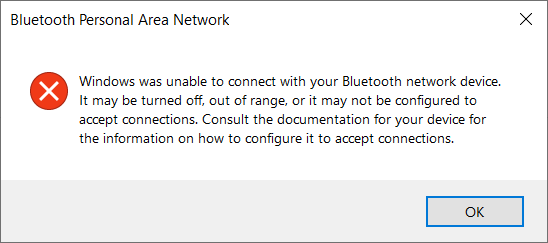}
    \caption{Windows error message when the tethering AP is off.}
    \label{fig:win_off}
\end{subfigure}
\caption{Error messages with an iPhone for tethering.}
\label{fig:errormessages}
\end{figure}

\subsection{Android (Samsung)}

Using a F\reflectbox{R}IDA-based \ac{PoC} on the \emph{Samsung Galaxy Note20 5G} on a
January 2021 patch level, we test key change behavior over-the-air against all devices.
The \emph{Samsung}-flavored user interface looks differently than the \emph{Google} user interface,
including menu structures and texts. When a \ac{BT} connection fails due to a changed link key, 
the message \emph{``Couldn't connect.''} is displayed. In case the paired device supports a special protocol or capability,
this is added to the text. For example, when using tethering with an \emph{iPhone}, the message
\emph{``Couldn't connect. Turn on Bluetooth tethering on iPhone.''} is shown, as depicted in \autoref{fig:sandroid_keychange}.
This \textbf{error message is the same as when the
paired device was switched off}, meaning that harmless connection issues and security-critical authentication errors are indistinguishable for users.

\emph{Google's Fast Pair} protocol substitutes pairing~\cite{fastpair}. Only a limited subset of devices
supports this protocol, such as the \emph{Bose QC35 II} headphones. 
\emph{Fast Pair}
boils down to setting a link key via \ac{HCI}. Thus, we can use the same \ac{PoC} to replace the key.
The internal logic of key management is the same, independent from \emph{Fast Pair}, and the user is displayed
the same \emph{``Couldn't connect.''} message.

On \emph{Android}, \textbf{\ac{BLE} devices} usually require using the according app by the vendor,
since \ac{BLE} is mostly used by IoT devices and has very diverse use cases. Vendor apps
do not provide insights on the current pairing state. Thus, we use the
\emph{nRF Connect} app~\cite{nrfconnect}, which allows connecting to \ac{BLE} services and control
the bonding state.
The \emph{nRF Connect} app uses the \emph{Android} Bluetooth API, meaning that some error messages are displayed
by the system upon errors. For example, when changing the \ac{BLE} long-term key for encryption during the initial pairing,
the message \emph{``Couldn't pair with MI Band 2. Make sure that it's ready to pair.''} is displayed.
When changing the encryption key later on in the \path{LE_Start_Encryption} command, the following
\emph{Encryption Mode Change} event indicates that the encryption could not be switched on
while using the wrong encryption key.
Thus, \emph{Android}
\textbf{terminates the connection but does not display any error message}.


\subsection{iOS}


We hook the Bluetooth daemon with a F\reflectbox{R}IDA-based \ac{PoC} on an \emph{iPhone 8} on \emph{iOS 14.4}.
After installing a Bluetooth debug profile, 
 we can use \emph{PacketLogger} to observe all \ac{HCI} packets. This provides
us with a powerful debug tool for the \emph{Apple} ecosystem, supporting devices like \emph{AirPods} that only
function with other \emph{Apple} devices.

When connecting to a \ac{BT} device, the error message is always \emph{``Connection Unsuccessful''}, as shown in \autoref{fig:ios_keychange}.
This error message is the same no matter if the other device is turned off or if the link key changed. \textbf{When switching back to the
original key, connections are successful again.}
The same error message is shown when porting the \ac{PoC} to a jailbroken \emph{iPhone 12} on \emph{iOS 14.1} and connecting
it to another \emph{iPhone 12} on \emph{iOS 14.6 Beta}, because the issue is anchored in the user interface and not the hardware.

\emph{Apple} uses \emph{MagicPairing} for \emph{AirPods}~\cite{magicpairing}.
It leverages the same mechanism as \emph{Fast Pair}---exchanging a cloud-based key 
and then setting it via an \ac{HCI} command.
When substituting the key in this command, the same error message is shown.

\emph{iOS} does not directly support third-party \ac{BLE} devices, 
and does not show them in the scan results. The \emph{nRF Connect} app for \emph{iOS} 
does not support bonding. Thus, we are not able to test non-\emph{Apple} \ac{BLE} devices on \emph{iOS} in a comparable fashion.

The \emph{Bose QC35 II} send \ac{BLE} advertisements and usually pair using \ac{BLE} followed by a cross-transport key derivation to
switch to \ac{BT}. However, after receiving the first \ac{BLE} advertisement from a \emph{Bose QC35 II},
\emph{iOS} requests further information via a \ac{BT} extended inquiry 
and directly pairs or connects using \ac{BT}.

\subsection{macOS}

We test the \emph{macOS} stack by connecting it to devices that switch their key.
The version under test is \emph{macOS 11.2.1} on a \emph{2020 MacBook Pro}.
No matter if connecting to a device via the menu bar on top or via the full settings dialogue, \textbf{buttons temporarily change their
color to blue}, similar to a connect and disconnect action, for \SIrange{2}{3}{\second}.
As on \emph{iOS}, this dialog does not support non-\emph{Apple} \ac{BLE} devices.

\subsection{Ubuntu with Gnome}

A default \emph{Ubuntu 20.10} installation uses \emph{Gnome} as user interface on top of the \emph{BlueZ} Bluetooth stack.
We use a \emph{ThinkPad X240} with these packets:
\texttt{gnome-control-center (1:3.38.3-0ubuntu1)}, \texttt{bluez} \texttt{(5.55-0ubuntu1.1)}, 
and \texttt{linux-kernel (5.8.0-44-generic)}. 
Instead of hooking into the Bluetooth daemon itself, we change keys within the file system.

For \ac{BT}, 
only button colors change to blue for a short moment.
We use the same \emph{ThinkPad X240} as for the \emph{Windows} setup, with the
only exception being the test between \emph{Linux} and \emph{Windows}, for which
we use a \emph{ThinkPad X1 Yoga} with \ac{BT} 4.2 as \emph{Linux} device.
When testing \ac{BLE} with the \emph{MI Band 2}, the initial pairing works, but even
without changing the key, reconnecting later on is not supported. Thus, we could not test the \ac{BLE}
behavior of this user interface, even though the underlying \emph{BlueZ} stack supports arbitrary
\ac{BLE} devices. 

\textbf{The \emph{BlueZ} stack is by far the most unreliable Bluetooth stack.} While testing the listed devices,
we observed one crash in the kernel module and two crashes in the Bluetooth daemon.

\subsection{Windows}

We use the most recent \emph{Windows 10 Internal Build} as available in March 2021.
\emph{Windows} has two menus that can connect to devices. First, the connect side bar is
reachable via \includegraphics[width=0.65em]{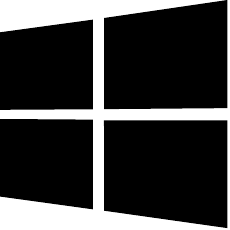} + K. This is primarily meant for
audio devices and other devices are only shown as connection information. The \emph{Linux} laptop
is detected as audio device.
Upon a key change, audio devices in this menu show the following message: \emph{``That didn't work. Make sure your Bluetooth device is still discoverable, then try again.''}

To connect
\emph{Windows} to one of the smartphones that change their key, we need to pair them via the
\emph{Settings} menu, go to \emph{Control Panel $\rightarrow$ Hardware and Sound $\rightarrow$ Devices and Printers},
right-click the paired smartphone, and connect to it using the \emph{AP} option.
When connecting to a smartphone with a changed key, the message \emph{``An unexpected error occurred. Please
contact your system administrator.''} is shown (see \autoref{fig:win_keychange}).
\textbf{The authentication failure error message is not 
helpful to determine the root cause} of not being able to connect to the smartphone.
Interestingly, \emph{Windows}
has the only interface where the message is different from the case of not being able to connect to 
switched off device (see \autoref{fig:win_off}). 

\subsection{Peripherals}
\label{ssec:vulnperipherals}

We test various types of peripherals: headphones, keyboards, and a \ac{BLE} fitness tracker.
\emph{AirPods} indicate connections with sounds, the \emph{Bose QC35 II} even reads out the
currently connected devices and pairing state, the \emph{Xiaomi MI Band 2} vibrates during the
initial pairing, the \emph{Mini Keyboard} indicates pairing states with an LED, and the
\emph{MagicKeyboard} lacks any kind of feedback mechanism.

Given these limited user interaction capabilities of peripherals, notifying the user of
a security failure, as suggested by the Bluetooth specification~\cite[p. 1314]{bt52}, requires special solutions.
While error sounds or status lights would be possible, \textbf{none of these devices indicate an error when using a wrong key}. When switching back to the correct
key, they accept the connection again.

\subsection{Outdated Bluetooth Versions}
\label{ssec:vulnoutdated}

Smartphones tend to have new Bluetooth chips supporting the most recent specification.
Yet, peripherals that require less throughput have surprisingly old chips. 
The device labeled as \emph{Mini Keyboard} is the cheapest keyboard in the \emph{Adafruit}
store sold for USD 12.95~\cite{minikeyboard}. It is using Bluetooth 2.1 + EDR, which has been
released in 2007. At least, this keyboard implements \emph{Passkey Entry} authentication.
Even popular recent devices such as standalone
\emph{MagicKeyboard} sold by \emph{Apple} in 2021 only has Bluetooth 3.0 + HS, dating
back to 2009. 
On \emph{Broadcom} chips, as used in this device, the firmware patching capabilities
are rather limited~\cite{frankenstein}. Issues that stem from the outdated Bluetooth
version in this chip cannot be fixed in software.
Most likely due to usability reasons, the \emph{MagicKeyboard} does not use
numeric verification during wireless pairing.

Keyboards are low-throughput, meaning that old chips do not have any noticeable effect for users.
However, security of keyboards is essential---users type confidential texts and passwords.
In addition to adding warnings on authentication failures as already required by the Bluetooth specification,
we suggest that \textbf{users should be warned about outdated Bluetooth versions.}

\subsection{Responsible Disclosure}
\label{ssec:disclosure}
We contacted the \emph{Bluetooth SIG}, \emph{Apple}, \emph{Google}, and \emph{Samsung}
on February 27th 2021. After building further \acp{PoC} and testing more devices, we 
contacted \emph{Microsoft}, \emph{Bose}, \emph{Xiaomi}, and \emph{Gnome} on March 13th.
The \emph{Bluetooth SIG} will address the issue. Moreover, \emph{Apple}, \emph{Google}, and
\emph{Samsung} will integrate warnings in a future release, but classified the issue as feature request.
\emph{Microsoft} stated that they will not change their warnings. \emph{Xiaomi} misunderstood the
report despite multiple clarifications. \emph{Bose} and \emph{Gnome} did not reply.


%% file: sections/conclusion.tex

\section{Conclusion}
\label{sec:conclusion}

While many researchers looked into cryptographic aspects of Bluetooth security,
little has been done to raise the bar for practical \ac{MitM} attacks.
Patching the newest cryptographic bugs within operating systems does not
structurally improve Bluetooth security, as peripherals remain outdated.
Users should be notified of security failures as proposed by the Bluetooth
specification. This would make the life of wireless attackers much harder, as it
significantly reduces attack stealthiness.
In addition, users should be warned if security-sensitive
peripherals like keyboards use a 10 year old Bluetooth version, vulnerable
to various known issues. On a long-term perspective, this would prevent vendors
from selling outdated peripherals.
Such structural improvements require everyone to contribute, admit flaws, and
indicate them towards the users---even if this might be inconvenient.